# Coupled Plasmonic-Waveguide Resonance Geometry for Enhanced Infrared Absorption in Semiconductor Solar Cells

Mohammad Abutoama,[1,2]

[1]Department of Electrical and Photonics Engineering, Technical University of Denmark, Ørsteds Plads, building 343, 2800 Kgs. Lyngby, Denmark

[2]NanoPhoton - Center for Nanonphotonics, Ørsteds Plads, building 345A, 2800 Kgs. Lyngby, Denmark

moh.abutoama@gmail.com

**Abstract**

Thin films are preferred for high photocurrent conversion efficiency, but strong photon absorption at photon energies below the bandgap (near- and shortwave infrared) typically requires thicker semiconductor layers. To address this trade-off, various optical approaches have been proposed, including light scattering within the active layer, reducing surface reflection, and using resonant structures to improve light confinement, trapping, and coupling. However, resonant structures often operate over a narrow spectral range, limiting their use of the full solar spectrum, and can involve complex fabrication and careful structural design. In this work, I propose a new method to enhance absorption in semiconductor solar cells across wide angular and spectral ranges for both polarization states (transverse electric (TE) and transverse magnetic (TM)). The method is based on a coupled plasmonic–waveguide resonance (CPWR) configuration excited in a planar layered structure that can be fabricated using simple deposition techniques. Using the proposed approach, as an example, the thickness of the required Silicon (Si) layer can be reduced from approximately 130–180 µm (the typical Si thickness in commercial solar cells) to only a few microns. The method enables efficient harvesting of the infrared portion of the solar spectrum. By exciting CPWRs, the method overcomes the sharp drop in the absorption spectrum of conventional Si solar cells at wavelengths longer than 1100 nm. Field calculations demonstrate that light is efficiently absorbed in the Si layer at the resonant wavelengths. The proposed approach is general and can be applied to different types of semiconducting and prism materials. To maintain high absorption at wavelengths below 1100nm, an additional semiconductor-metal-semiconductor configuration is proposed, in which a thinner Si layer is added beneath the metal layer.

**Motivation**

Solar energy is considered one of the most important energy sources for the future. Therefore, proposing solutions to enhance photoenergy conversion efficiency is crucial to the development of advanced energy devices, including photovoltaics, photocatalysis, solar vapor generation, photoelectrochemical water splitting, water desalination, and light-emitting devices. Low optical absorption and high electron–hole recombination are among the major challenges that significantly limit the performance of these energy devices. For example, there is an inherent trade-off between minimizing electron–hole recombination, which favors thinner layers, and maximizing photon absorption for improved photocurrent generation, which requires thicker layers in conventional solar cells.

Silicon (Si) solar cells typically absorb light in the ~400–1100 nm spectral range. As an indirect bandgap semiconductor ($E_g \approx 1.12$ eV), the absorption of Si drops sharply near ~1100–1200 nm. The thickness of typical commercial Si solar cells is about ~130–180 µm. Thin-film Si technology is a promising candidate for addressing the significant increase in global energy consumption. Reducing the thickness of Si solar cells enables higher industrial throughput while decreasing material usage and production costs. However, maintaining or improving the conversion efficiency of thinner solar cells requires strategies to increase the effective optical path length of light within the active layer to enhance absorption. To better understand these strategies, they can be broadly divided into two main categories: resonant and non-resonant techniques.

The basic idea behind conventional (non-resonant) absorption enhancement techniques is either to reduce reflection from the solar cell or to increase light scattering within the active material. For example, antireflection coatings (ARCs), typically composed of thin dielectric layers, can increase the probability of photon absorption in the active semiconductor layer. Surface texturing methods can also be employed to create sharp features, such as pyramid-like structures, on the Si top surface to scatter and trap incident light, thereby reducing reflection and increasing the optical path length of light within the solar cell [1]. Surface texturing can be implemented in various geometries, such as nanocones [2], and can even involve randomly shaped structures. Inspired by the apposition compound eyes of certain dipterans, the use of prismatic lenses with specific cross-sectional shapes was investigated in [3]. The results suggest that texturing the exposed surfaces of Si solar cells into arrays of bioinspired compound lenses improves performance.

On the other hand, optical resonators have emerged as one of the most promising approaches for enhancing the performance of solar cells. Resonant techniques rely on advanced optical nanostructures that employ various mechanisms to enhance absorption in thin-film solar cells, such as light trapping and scattering, and to enable efficient coupling of incident light into the active layer. However, a major limitation of resonant structures is that they typically operate at specific wavelengths, leaving a large portion of the solar spectrum unutilized. Therefore, there is a strong need for alternative and advanced strategies or structures that, first, exhibit broadband absorption [4] across the wavelength range of 200–2500 nm, corresponding to the solar irradiation spectrum at sea level, and, second, generate strongly enhanced electromagnetic fields to promote efficient generation and separation of electron–hole pairs while reducing electron–hole recombination.

**Applications of micro and nanoscale optical resonators in Photovoltaics**

The use of optical cavities in photovoltaic applications has been discussed in a limited number of reviews [5–6]. The fundamentals and recent progress in optical-resonator-enhanced photovoltaics were comprehensively presented and discussed in [7]. This review reports both theoretical and experimental advances in photovoltaic devices by exploiting the unique features of microscale and nanoscale optical resonators, such as strong light confinement and enhanced light–matter interactions. These resonators include Fabry–Perot (FP) cavities [8], whispering-gallery-mode (WGM) cavities [9], photonic crystal (PC) cavities [10], plasmonic resonators [11], and hybrid cavities. Optical-resonator-based photovoltaics have the potential to overcome the trade-off between light absorption and carrier diffusion [7].

The first thin-film photovoltaic (PV) cells based on an FP cavity with dielectric mirrors were demonstrated in 2008 [8] to enhance optical absorption. In this structure, the active layer was sandwiched between the mirrors of the FP cavity. An important observation is that this configuration resulted in a decreased response at shorter wavelengths and an increased response at longer wavelengths, better matching the solar spectrum [8]. Similar work was also reported in [12]. Different types of FP-cavity-based solar cells have been proposed. For instance, organic materials can be placed within an FP cavity to achieve strong coupling between excitons and the optical cavity mode [13], resulting in a splitting of the absorption spectrum of the active material into two polariton peaks. The integration of quantum dots within an FP-cavity-based solar cell was demonstrated in [14], where the resonant absorption can be tuned by adjusting the thickness of the active layer. In general, the mechanism responsible for performance enhancement in FP-cavity-based solar cells involves increasing optical absorption in the active layer sandwiched between the two mirrors of the cavity and enabling manipulation of the electromagnetic field within the active layer [7].

WGM cavities have also been utilized in solar cells and photovoltaic applications [9, 15–17]. These include the implementation of WGM resonances in dye-sensitized solar cells [18], quantum-dot-sensitized solar cells [19], and high-efficiency perovskite solar cells, in which incident light can be confined within the perovskite layer via total internal reflection [20]. Such structures enhance light trapping, promote electron–hole separation, and significantly reduce carrier recombination [20]. In general, the performance enhancement in WGM-cavity-based solar cells arises from improved light confinement and increased optical scattering within the active medium [7].

PC-cavity-based solar cells consist of periodic structures in one, two, or three dimensions [10, 21–29]. Generally, the enhanced performance of PC-cavity-based solar cells relies on improved light trapping within the active layer, increased local density of states (LDOS) at the band edge, and the use of the "slow-photon" effect to extend the optical path length [7]. Periodic structures, such as PCs and gratings, effectively couple incident light and significantly increase the optical path length of photons within the semiconductor layer [30–31].

Plasmonic-cavity-based solar cells can be designed using various metallic structures, such as nanoparticles with dimensions much smaller than the incident-light wavelength to support localized surface plasmon resonance (LSPR), metallic gratings, and thin continuous metal films that support propagating surface plasmons (SPs). Plasmonic nanoparticles have been widely employed to improve solar cell performance [11, 32–34]. These nanoparticles can be placed on the surface or embedded within thin-film solar cells to efficiently trap and scatter light via LSPR excitation, thereby enhancing near-field intensity. Broadband absorption enhancement has been achieved by combining nanoparticles with gratings [35]. Gap plasmons can also be used to efficiently couple incident light with the semiconductor layer [36], resulting in remarkable absorption enhancement [37]. Bowtie structures can act as light collectors and scatterers [38–39], enabling performance improvements over a wide spectral range.

However, individual metallic nanoparticles typically resonate over a narrow wavelength range, limiting efficient utilization of the solar spectrum. In [40], a rational design of plasmonic nanostructures for solar energy conversion was proposed using multiplexed nanoparticle configurations (varying composition, size, shape, or array arrangement) to provide a versatile platform for enhancing absorption and maximizing light-trapping efficiency.

The mechanism behind performance enhancement in plasmonic-cavity-based solar cells relies on increased absorption, efficient light scattering within the photovoltaic layer, improved carrier collection, and broadening of the absorption spectral width [7]. It should be noted that plasmonic effects do not always improve solar cell efficiency due to quenching [7]. To mitigate this issue, plasmonic structures should be physically separated from the active material.

Hybrid-cavity-based solar cells hold significant potential for designing unique, high-efficiency devices in which two or more different optical modes can be coupled to achieve a strong-coupling regime [7]. These structures combine the advantages of multiple cavity types, further enhancing light-harvesting capabilities. Several examples of hybrid cavities have been reported: strong coupling between plasmonic and FP modes has been demonstrated, resulting in field enhancement within the absorption band [41]; strong coupling between FP and LSPR modes has been used to realize compact photovoltaic devices [42]. Additionally, overall absorption enhancement has been proposed based on the coupling between Mie resonances excited in dielectric structures and FP cavity modes [43–44].

An additional resonant approach for designing advanced solar cells is the use of tapered silicon nanowire (SiNW) arrays, which enable broadband, high-efficiency absorption without the need for thick Si layers [45].

As an extension of the propagating SP geometry, a modified design in which a dielectric layer is inserted on top of a thin metallic film has been reported [46–49]. In this configuration, known as the coupled plasmon–waveguide resonance (CPWR) [50], the dielectric layer acts as a waveguide, supporting guided modes in addition to SP modes. Unlike conventional propagating SP schemes, the CPWR system can excite guided modes in both TM and TE polarization states [50].

Building on the fact that guided modes are excited within the dielectric top layer, we propose replacing the dielectric layer with a semiconductor material, specifically Si. Due to the large propagation distances at the analyte interface, the guided modes are highly efficient candidates for sensing applications [50]. To the best of our knowledge, no studies have reported the use of the CPWR configuration in energy-related devices, particularly for enhancing absorption in Si solar cells. This constitutes the main focus of the current work.

Despite the performance enhancements achieved in optical-resonator-based solar cells, several challenges may limit their practical implementation. For example, FP–based solar cells require precise mirror alignment and are often sensitive to the angle of incidence. In WGM-based solar cells, the overlap between the active material and the enhanced electromagnetic fields can be low. Plasmonic systems may suffer from quenching effects, and hybrid-cavity-based solar cells require careful optimization of structural parameters as well as relatively complex fabrication processes.

In this work, we report an additional mechanism for enhancing absorption in Si solar cells without requiring thick active layers. It has been shown that the thickness of a commercial Si solar cell can be reduced from ~150 µm to just a few micrometers. This approach is not limited to Si and can be applied to other semiconductor materials. The proposed geometry consists of planar films that can be fabricated using standard deposition methods, making it convenient for both modeling and optical calculations, and avoiding the complex fabrication requirements of hybrid structures. The effectiveness of the design is demonstrated over a wide range of incident angles and wavelengths, for both TE and TM polarization states.

The structure under study consists of a coupling medium made of high-index prism (SF11 glass), coated with a thin silver film (Ag), and covered with a waveguide layer made of silicon (Si), as shown in Figure

1(a). The thickness of the Ag film is fixed throughout the study at $d_{Ag}$ = 45 nm. The incidence angle is denoted by θ, as presented in Figure 1(a).

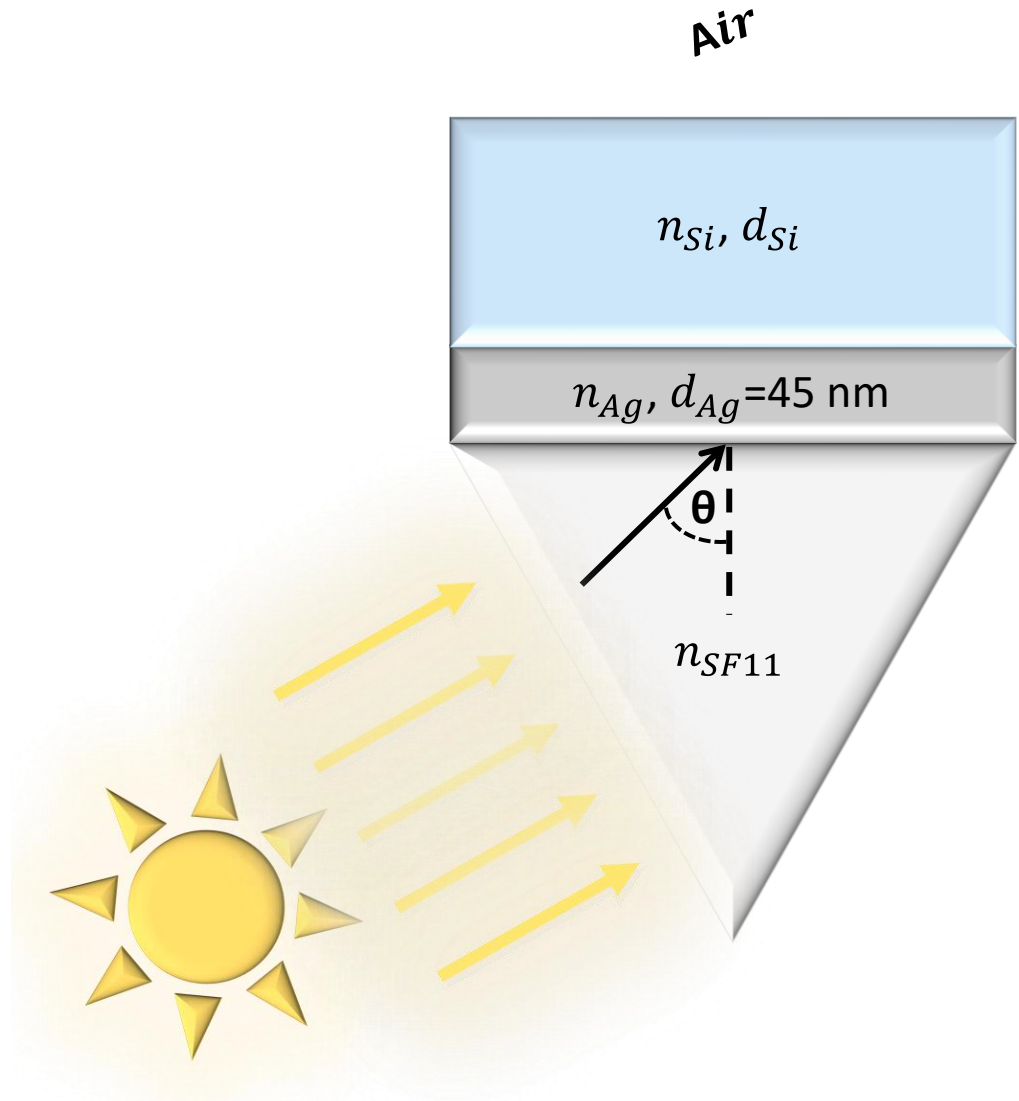

Fig.1. Schematic of the CPWR-based Si solar cell. The Ag and Si layers are bounded by two semi-infinite dielectric media: the ambient and the substrate materials (air and SF11, respectively, in our case).

## Results and Discussion

Computational studies

To simulate the reflectivity and field distribution of the structure shown in Figure 1(a), we used the Shalabney-Abdulhalim algorithm based on the Abeles matrices [51]. The dielectric functions of the materials were interpolated from data obtained from the Sopra database (http://sspectra.com/). The code was implemented using MATLAB.

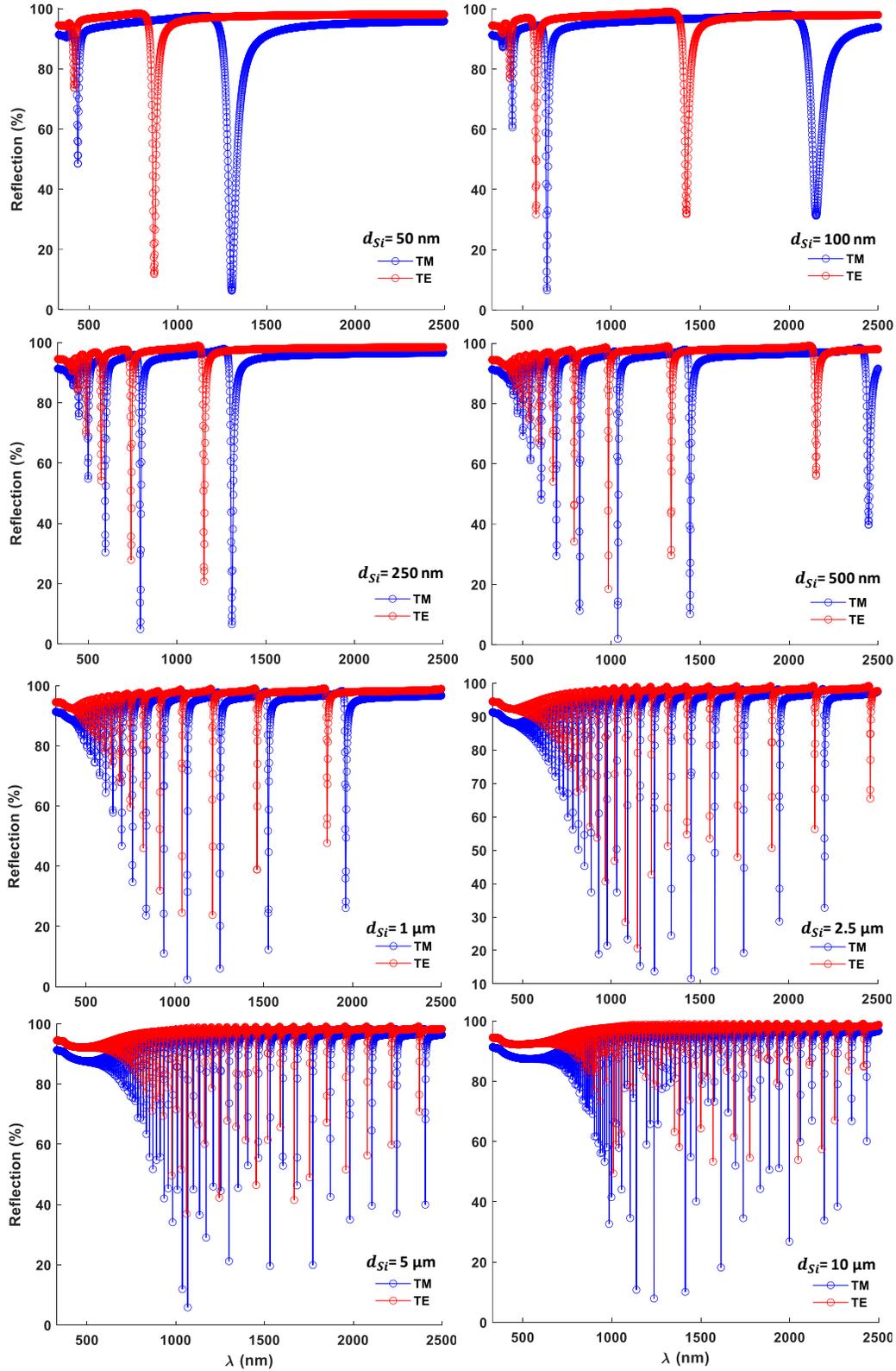

Figure 2. Reflection spectra for the structure shown in Figure 1 at θ =45º and $d$Ag = 45 nm, for the two polarization states (TE and TM), and for different thicknesses of the Si layer.

The concept behind the proposed geometry and its ultimate goal is to obtain a structure that exhibits a large number of resonances to increase photocurrent generation and, more importantly, to distribute these resonances across a wide spectral range for better utilization of the solar spectrum. To demonstrate this approach, we start with thin Si layers on top of the Ag layer and gradually increase their thickness from 50 nm to 10 µm, as shown in Figure 2. The simulations were performed at a fixed incidence angle of θ = 45º for both polarization states, transverse electric (TE) and transverse magnetic (TM), over the spectral range of 330–2500 nm. For $d_{Si}$ = 50 nm, two resonances can be observed for TE and TM polarizations in the top-left part of Figure 2. As $d_{Si}$ increases, more resonances appear for both polarizations. It can be seen that for $d_{Si}$ larger than 5 µm (the two bottom subfigures in Figure 2), a large number of resonances are observed, spanning approximately 600–2500 nm.

The absorption enhancement is shown in Figure 3 for TE and TM polarization at θ = 45 ° and $d_{Si}$ = 10 µm. The blue curves represent the absorption spectrum without the 45 nm Ag thin film, while the black curves correspond to the case with the Ag thin film. As discussed in the introduction, the absorption of Si (without Ag) drops significantly at wavelengths larger than approximately 1100 nm. The inclusion of the Ag thin film excites CPWRs in the structure, which in turn introduces numerous resonances, particularly at wavelengths longer than approximately 1100 nm.

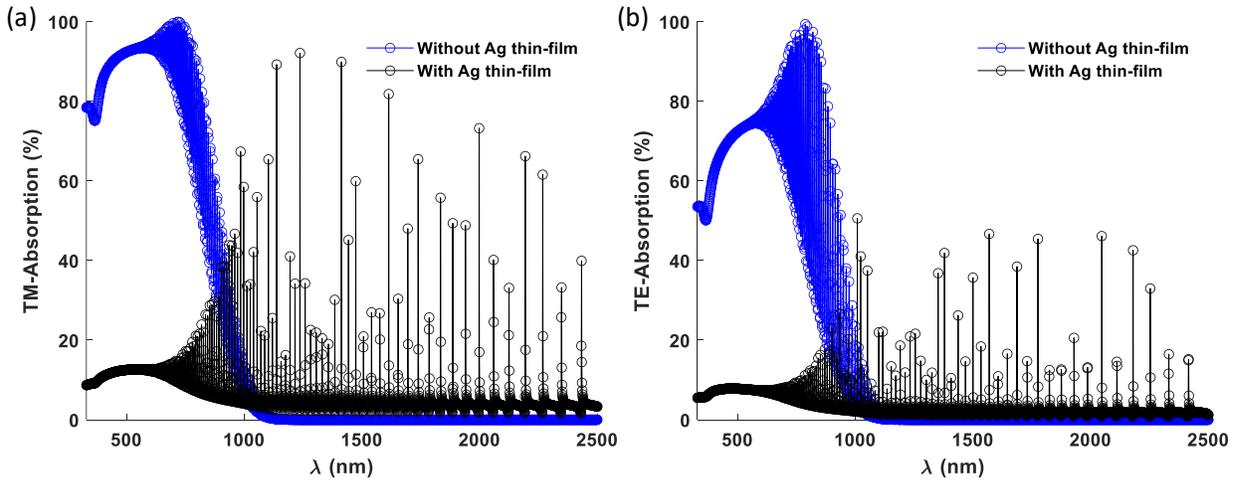

Figure 3. Absorption spectra of the structure shown in Figure 1 at θ =45º, $d_{Ag}$ = 45 nm, and $d_{Si}$ = 10 µm, for the two polarization states (TE and TM).

Without the Ag layer, CPWR excitation is not possible. Figure 3 demonstrates the potential of the CPWR-based solar cell geometry to enhance absorption in the semiconductor layer in the infrared regime.

Next, it is important to verify that the electromagnetic field is confined within the Si layer at the resonant wavelengths. As an example, we consider $d_{Si}$ = 250 nm, which exhibits resonances at 417, 448, 500, 597, 794, and 1310 nm, as shown in Figure 4.

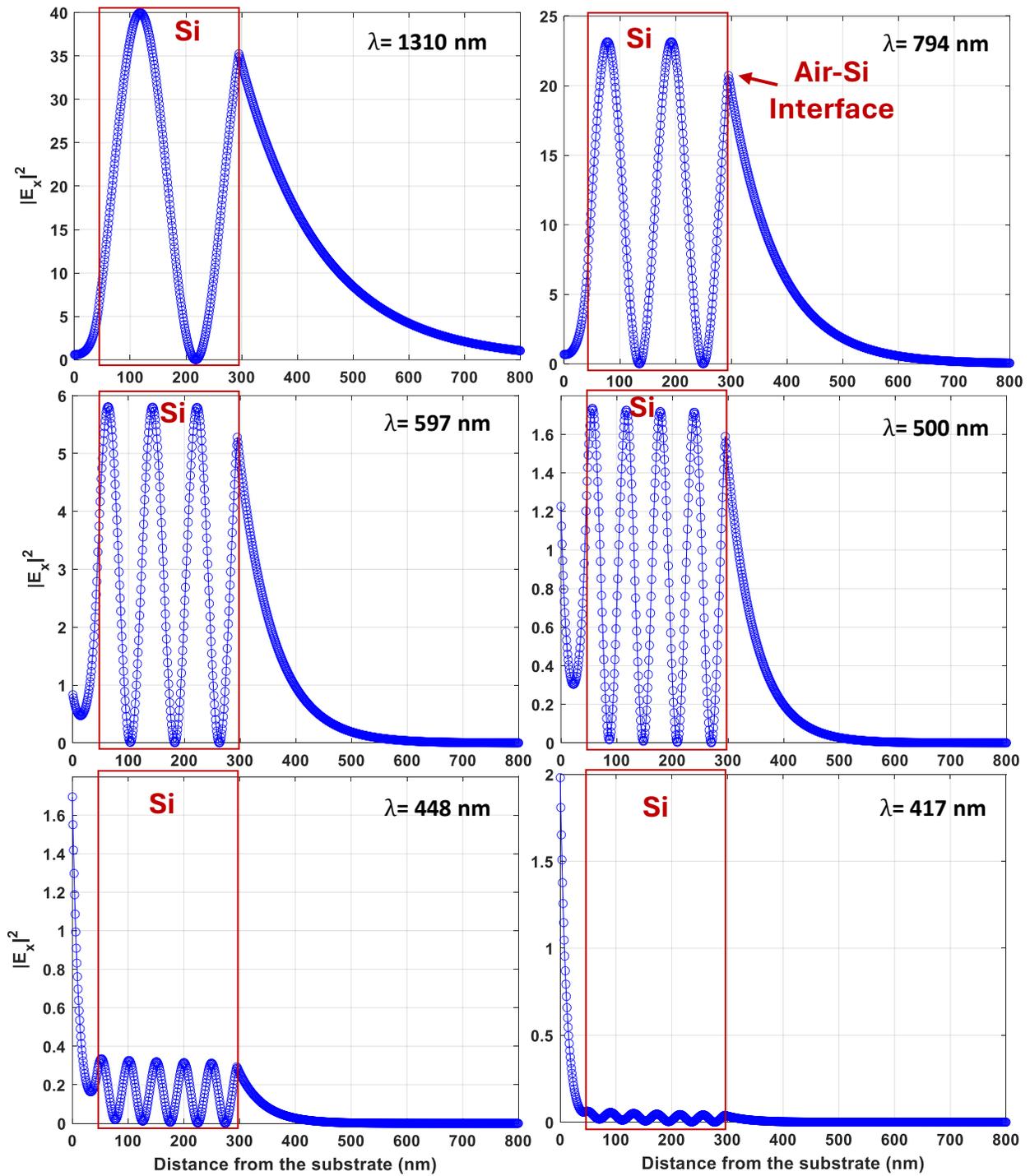

Figure 4. Field distribution at the resonant wavelengths (intensity of the x-component of the electric field) in the CPWRs at θ =45°, $d_{Ag}$ = 45 nm, $d_{Si}$ = 250 nm, for TM polarization.

The field is indeed present and dominant within the Si layer at the resonant wavelengths, which explains the absorption enhancement observed at those wavelengths in Figure 3. In the air region, the field is evanescent, as expected.

To determine the fraction of light absorbed in the waveguide (Si) compared to that in the metal, we calculate the total energy by integrating the field intensity in the waveguide and dividing it by the integrated field intensity in the metal. This ratio is expressed as $R$:

$$(1) \qquad R = \frac{\varepsilon_{Si} \int_{d_{Ag}}^{d_{Si}} |E_{Si}(z)|^2 dz}{\varepsilon_{Ag} \int_0^{d_{Ag}} |E_{Ag}(z)|^2 dz}$$

As an example, we calculate $R$ for the same configuration shown in Figure 4 ($\theta = 45°$, $d_{Ag} = 45$ nm, $d_{Si} = 250$ nm, for TM polarization). We obtained $R$ values of approximately 5.41, 9.28, and 0.09 at the three wavelengths $\lambda = 448$, 1310, and 1700 nm, respectively. As expected, these values are consistent with the resonant behavior: 448 nm and 1310 nm correspond to resonant wavelengths, as shown in Figure 4, while $\lambda = 1700$ nm is a non-resonant wavelength. This confirms that the enhanced absorption at the resonances is indeed due to absorption in the Si.

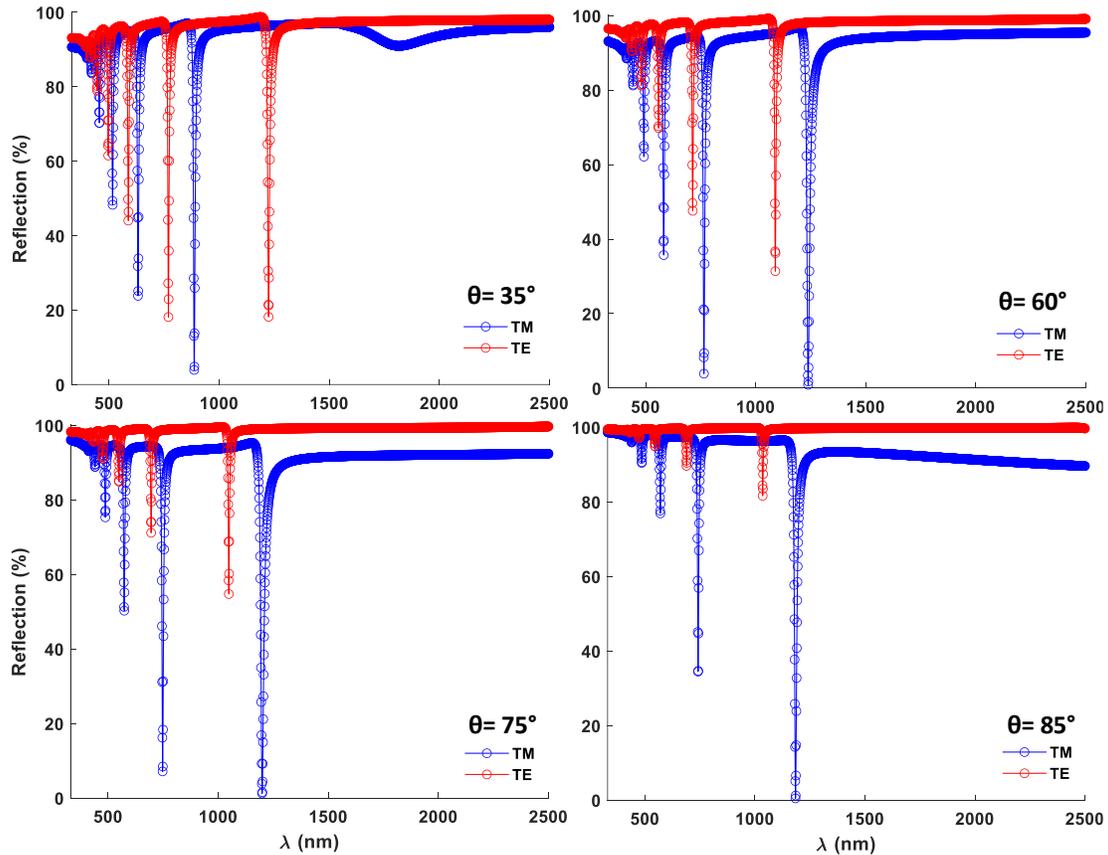

Figure 5. Reflection spectra of the structure shown in Figure 1 at different incidence angles (35°, 60°, 75°, and 85°), with $d_{Ag} = 45$ nm and $d_{Si} = 250$ nm, for the two polarization states (TE and TM).

In all the simulations above, a fixed incidence angle of $\theta = 45°$ was used. However, an efficient energy device should exhibit good performance over both a wide spectral range and a wide angular range. In Figure 5, we investigate the response of the CPWR-based solar cell system at different incidence angles of 35°,

60°, 75°, and 85°. The structure parameters are $d_{Ag}$ = 45 nm, $d_{Si}$ = 250 nm, and the analysis is performed for both TE and TM polarizations.

It is encouraging to observe that the structure remains resonant across a wide angular range. The key requirement is that the incident light angle must exceed the critical angle to excite the CPWRs, which in this case is approximately 34.5º.

### Effect of the prism material

Although high-index prisms have a lower critical angle and increased absorption, it is important to check whether off-the-shelf, low-cost prisms or prism arrays, such as those made of silica or BK7 glass, are available. The absorption enhancement is shown in Figure 6 for TE and TM polarization at θ = 45º, with $d_{Si}$ = 10 µm, using a BK7 prism instead of an SF11 prism. As shown in Figure 5, the structure remains resonant over a wide angular range; the critical angle in this case is approximately 41.5º. The blue curves represent the absorption spectrum without the 45 nm Ag thin film, while the black curves correspond to the case with the Ag thin film.

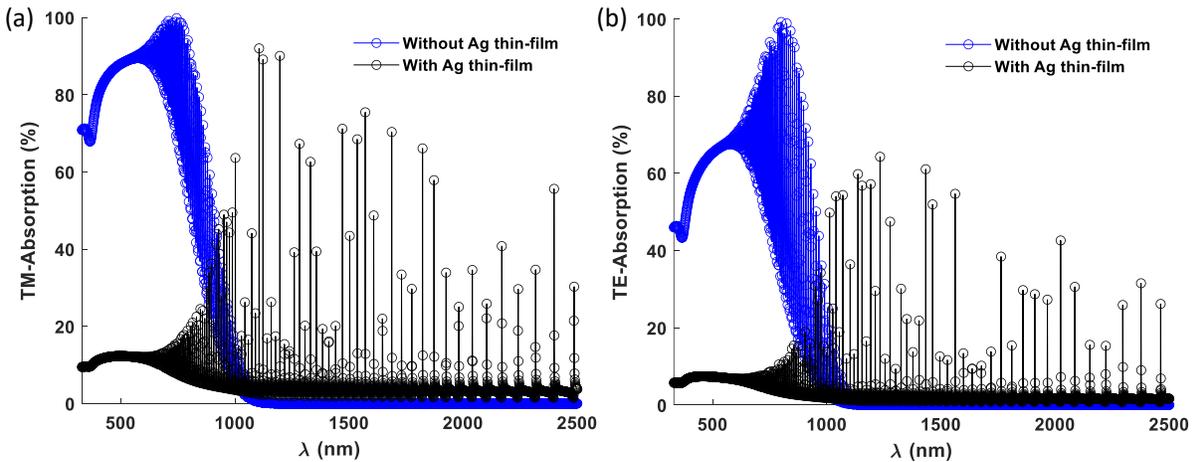

Figure 6. Absorption spectra of the structure shown in Figure 1, but with a BK7 prism instead of SF11 prism, at θ =45º, $d_{Ag}$ = 45 nm, and $d_{Si}$ = 10 µm, for the two polarization states (TE and TM).

### Semiconductor-Metal-Semiconductor geometry to enhance absorption at short wavelengths

The proposed geometry in Figure 1 results in a significant enhancement of absorption in the infrared range (the black curves in Figures 3 and 6). However, this comes at the cost of reduced absorption at short wavelengths compared with the case without the Ag layer (the blue curves in Figures 3 and 6), due to the enhanced reflectivity at the silver interface.

To address this issue, a modified CPWR-based Si solar cell geometry is proposed in Figure 7 to enhance absorption at short wavelengths. The structure is similar to that shown in Figure 1, except for the addition of a 1 μm-thick Si layer between the prism and the Ag layer. We call it a semiconductor-metal-semiconductor (SMS) structure.

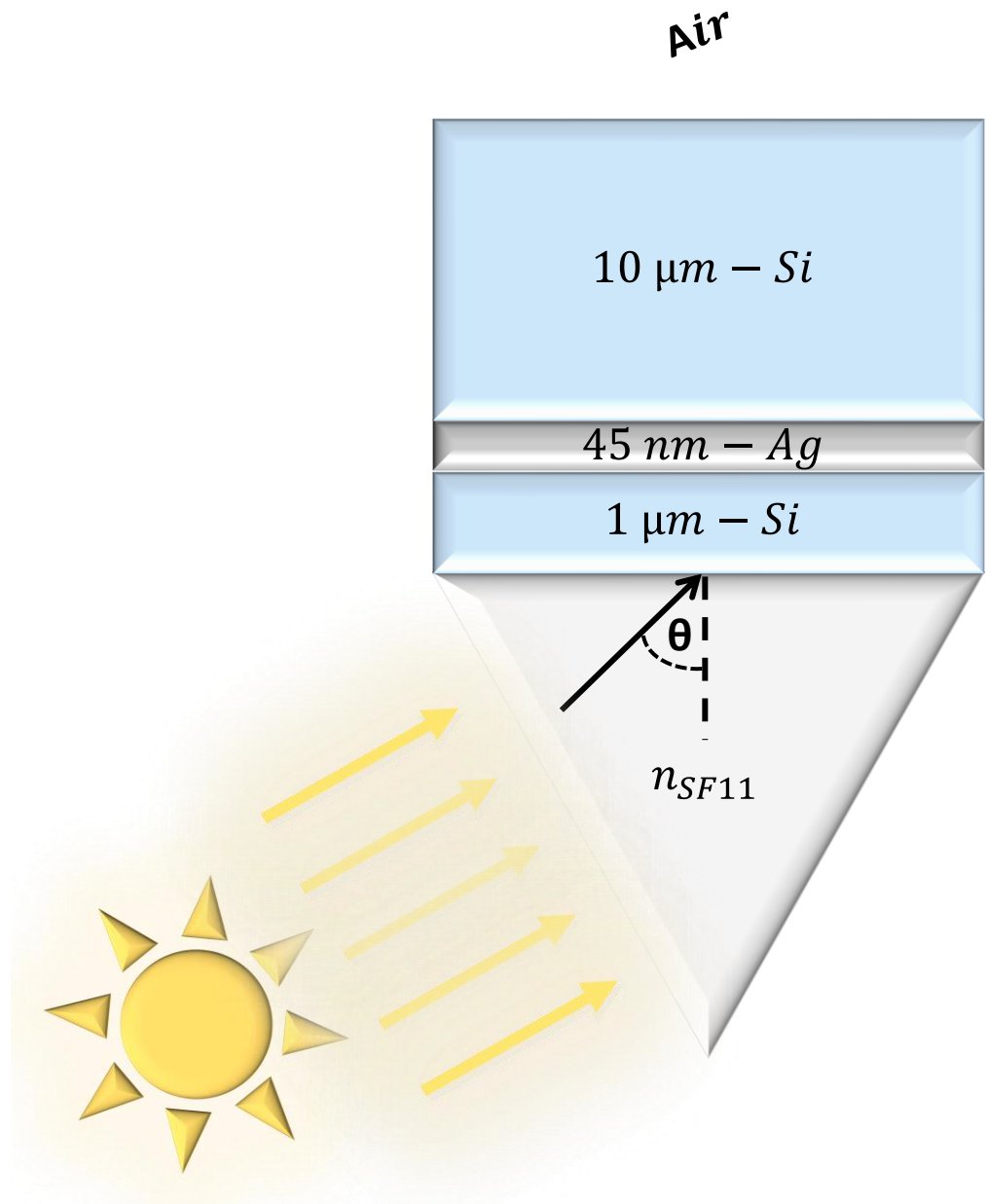

Figure 7. Schematic of the modified CPWR-based Si solar cell called SMS geometry. The structure is similar to that shown in Figure 1, except for the addition of a 1 μm Si layer between the SF11 prism and the Ag layer.

The reflection spectra for the structure in Figure 7 are shown in Figure 8. The simulation was performed at θ = 45º for both polarization states, TE and TM. A large number of resonances are observed, spanning approximately 400–2500 nm.

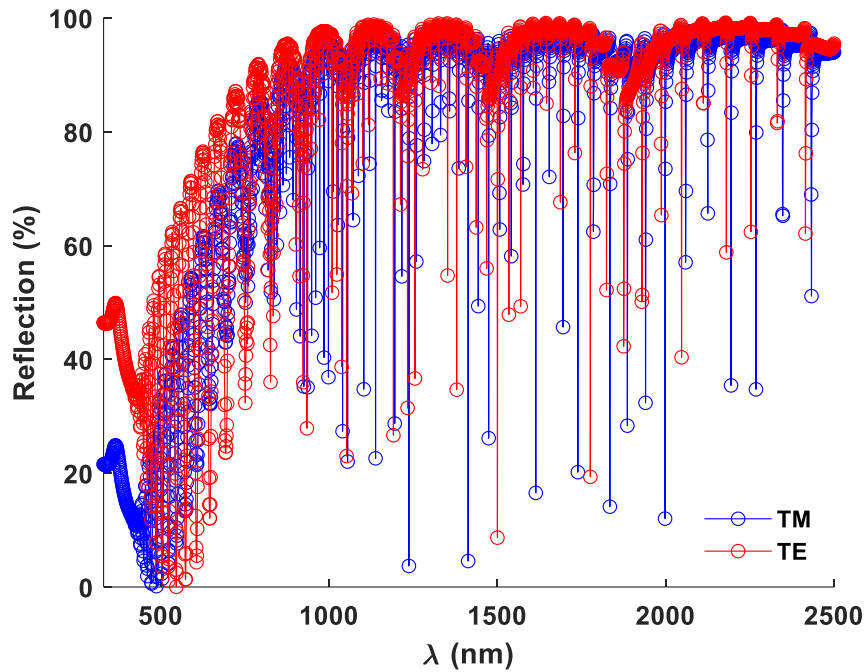

Figure 8. Reflection spectra of the modified structure shown in Figure 7 at θ =45°, for the two polarization states (TE and TM).

The absorption enhancement for the modified structure shown in Figure 8 is presented in Figure 9 for (a) TM and (b) TE polarization states at θ = 45°. The addition of the 1 μm Si layer also improves absorption at short wavelengths. Consequently, the modified CPWR-based Si solar cell geometry enhances absorption across almost the entire solar spectrum.

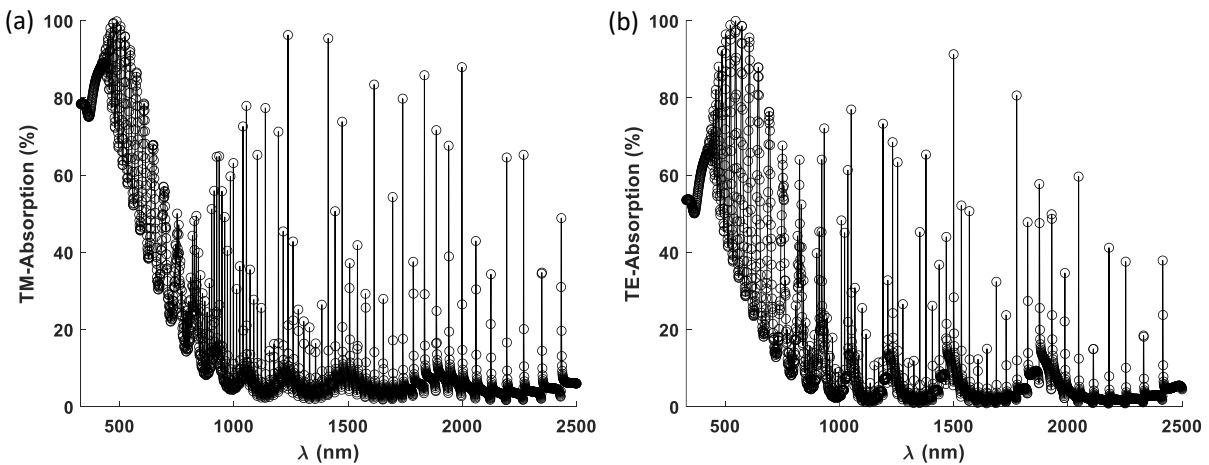

Figure 9. Absorption spectra of the SMS structure shown in Figure 7 at θ =45º, for the two polarization states: (a) TM and (b) TE.

Therefore, the CPWR geometry and, in particular, the SMS one can be understood as an extension of the existing Si solar cell to harvest the infrared spectrum.

**Conclusions**

A new method for enhancing absorption in semiconductor films is proposed based on the CPWR geometry. The structure combines a thin metal film supporting propagating SP modes with a waveguide film supporting guided modes for both polarization states (TE and TM). For further improvement of the solar cell performance, an additional Si layer is inserted between the prism and the metal layer to enhance absorption at short wavelengths in what we call SMS geometry. The uniqueness of the proposed geometry lies in its relatively simple fabrication process and its ability to operate over both a wide angular range and a wide spectral range without requiring thick semiconductor films. Using the proposed approach, the thickness of the required Si layer can be reduced from approximately 130–180 µm (the typical Si thickness in commercial solar cells) to only a few microns. Due to the excitation of CPWRs, the method overcomes the sharp drop in the absorption spectrum of conventional Si solar cells at wavelengths larger than approximately 1100 nm. As an example, application, Si films were used in this study; however, the proposed method is not limited to Si and can be applied to other semiconducting materials, such as perovskites, which have low absorption in the near infrared. In practice, there are two ways to implement the method over a large area: using a microprism array or, to achieve a similar effect, employing grating-coupling geometry instead of prism coupling. These extensions are planned for future work. This approach has great potential to improve photovoltaic efficiency, enhance light harvesting, reduce the cost of Si solar cells, and lower semiconductor material consumption.

**Declaration of competing interest**

The author declares no known competing financial interests or personal relationships that could have appeared to influence the work reported in this paper.

**References**

[1] L. Zhao, Y. H. Zuo, C. L. Zhou, H. L. Li, H. W. Diao, and W. J. Wang, *Sol. Energy*, 2011, **85**, 530.
[2] K. X. Wang, Z. Yu, V. Liu, Y. Cui, and S. Fan, *Nano Lett.*, 2012, **12**(3), 1616.
[3] F. Chiadini, V. Fiumara, A. Scaglione, and A. Lakhtakia, *J. Photonics Energy*, 2013, **3**, 034599.
[4] M. Abutoama, S. Li, and I. Abdulhalim, *J. Phys. Chem. C*, 2017, **121**, 27612.
[5] Y. Wang, P. Shen, J. Liu, Y. Xue, Y. Wang, M. Yao, and L. Shen, *Sol. RRL*, 2019, **3**, 1900181.
[6] W. Wang and L. Qi, *Adv. Funct. Mater.*, 2019, **29**, 1807275.
[7] Z. Yuan, P. C. Wu, and Y.-C. Chen, *Laser Photonics Rev.*, 2022, **16**, 2100202.
[8] M. Agrawal and P. Peumans, *Opt. Express*, 2008, **16**, 5385.
[9] J. Grandidier, D. M. Callahan, J. N. Munday, and H. A. Atwater, *Adv. Mater.*, 2011, **23**, 1272.
[10] J. Wang, Y. Yuan, H. Zhu, T. Cai, Y. Fang, and O. Chen, *Nano Energy*, 2020, **67**, 104217.
[11] J.-L. Wu, F.-C. Chen, Y.-S. Hsiao, F.-C. Chien, P. Chen, C.-H. Kuo, M. H. Huang, and C.-S. Hsu, *ACS Nano*, 2011, **5**, 959.
[12] Q. Liu, J. Toudert, T. Li, M. Kramarenko, G. Martínez-Denegri, L. Ciammaruchi, X. Zhan, and J. Martorell, *Adv. Energy Mater.*, 2019, **9**, 1900463.
[13] A. Kavokin and G. Malpuech, *Cavity Polaritons*, Elsevier, New York, 2003.

[14] O. Ouellette, N. Hossain, B. R. Sutherland, A. Kiani, F. P. García de Arquer, H. Tan, M. Chaker, S. Hoogland, and E. H. Sargent, *ACS Energy Lett.*, 2016, **1**, 852.
[15] J. Yin, Y. Zang, C. Yue, X. He, J. Li, Z. Wu, and Y. Fang, *Phys. Chem. Chem. Phys.*, 2013, **15**, 16874.
[16] G. Yin, P. Manley, and M. Schmid, *Sol. Energy Mater. Sol. Cells*, 2016, **153**, 124.
[17] A. Dabirian and N. Taghavinia, *RSC Adv.*, 2013, **3**, 25417.
[18] T. K. Das, P. Ilaiyaraja, and C. Sudakar, *ACS Appl. Energy Mater.*, 2018, **1**, 765.
[19] T. K. Das, P. Ilaiyaraja, and C. Sudakar, *Sci. Rep.*, 2018, **8**, 9709.
[20] Y. Wang, M. Li, X. Zhou, P. Li, X. Hu, and Y. Song, *Nano Energy*, 2018, **51**, 556.
[21] S.-J. Ha, J.-H. Kang, D. H. Choi, S. K. Nam, E. Reichmanis, and J. H. Moon, *ACS Photonics*, 2018, **5**, 3621.
[22] K. Kim, S. K. Nam, J. Cho, and J. H. Moon, *Nanoscale*, 2020, **12**, 12426.
[23] K. Kim, S. K. Nam, and J. H. Moon, *ACS Appl. Energy Mater.*, 2020, **3**, 5277.
[24] M. Portnoi, T. J. Macdonald, C. Sol, T. S. Robbins, T. Li, J. Schläfer, S. Guldin, I. P. Parkin, and I. Papakonstantinou, *Nano Energy*, 2020, **70**, 104507.
[25] H. C. Bauser, C. R. Bukowsky, M. Phelan, W. Weigand, D. R. Needell, Z. C. Holman, and H. A. Atwater, *ACS Photonics*, 2020, **7**, 2122.
[26] D. H. Choi, S. K. Nam, K. Jung, and J. H. Moon, *Nano Energy*, 2019, **56**, 365.
[27] L. Zhang, M. T. Hörantner, W. Zhang, Q. Yan, and H. J. Snaith, *Sol. Energy Mater. Sol. Cells*, 2017, **160**, 193.
[28] S. Baek, S.-J. Ha, H. Lee, K. Kim, D. Kim, and J. H. Moon, *ACS Appl. Mater. Interfaces*, 2017, **9**, 37006.
[29] M. Guo, Z. Yong, K. Xie, J. Lin, Y. Wang, and H. Huang, *ACS Appl. Mater. Interfaces*, 2013, **5**, 13022.
[30] Y. Park, E. Drouard, O. El-Daif, X. Letartre, P. Viktorovitch, A. Fave, A. Kaminski, M. Lemiti, and C. Seassal, *Opt. Express*, 2009, **17**, 14312.
[31] M. Eskandari and A. Habibzadeh-Sharif, *Photon. Nanostruct.: Fundam. Appl.*, 2024, **58**, 101229.
[32] D. Schaadt, B. Feng, and E. Yu, *Appl. Phys. Lett.*, 2005, **86**, 063106.
[33] D. H. Wang, D. Y. Kim, K. W. Choi, J. H. Seo, S. H. Im, J. H. Park, O. O. Park, and A. J. Heeger, *Angew. Chem. Int. Ed.*, 2011, **123**, 5633.
[34] M. Yao, X. Jia, Y. Liu, W. Guo, L. Shen, and S. Ruan, *ACS Appl. Mater. Interfaces*, 2015, **7**, 18866.
[35] X. Li, X. Ren, F. Xie, Y. Zhang, T. Xu, B. Wei, and W. C. Choy, *Adv. Opt. Mater.*, 2015, **3**, 1220.
[36] A. Ghobadi, T. G. Ulusoy Ghobadi, F. Karadas, and E. Ozbay, *Adv. Opt. Mater.*, 2019, **7**, 1900028.
[37] M. Choi, G. Kang, D. Shin, N. Barange, C.-W. Lee, D.-H. Ko, and K. Kim, *ACS Appl. Mater. Interfaces*, 2016, **8**, 12997.
[38] P. M. Voroshilov, V. Ovchinnikov, A. Papadimitratos, A. A. Zakhidov, and C. R. Simovski, *ACS Photonics*, 2018, **5**, 1767.
[39] P. M. Voroshilov, C. R. Simovski, and P. A. Belov, *J. Mod. Opt.*, 2014, **61**, 1743.
[40] Y. Wang, J. Zhang, W. Liang, H. Yang, T. Guan, B. Zhao, Y. Sun, L. Chi, and L. Jiang, *CCS Chem.*, 2022, **4**, 1153.
[41] Y. Jin, J. Feng, M. Xu, X. L. Zhang, L. Wang, Q. D. Chen, H. Y. Wang, and H. B. Sun, *Adv. Opt. Mater.*, 2013, **1**, 809.
[42] C. Hägglund, G. Zeltzer, R. Ruiz, A. Wangperawong, K. E. Roelofs, and S. F. Bent, *ACS Photonics*, 2016, **3**, 456.
[43] B. C. Sturmberg, T. K. Chong, D.-Y. Choi, T. P. White, L. C. Botten, K. B. Dossou, C. G. Poulton, K. R. Catchpole, R. C. McPhedran, and C. M. de Sterke, *Optica*, 2016, **3**, 556.
[44] V. K. Narasimhan, T. M. Hymel, R. A. Lai, and Y. Cui, *ACS Nano*, 2015, **9**, 10590.
[45] S. F. Sajadian, M. R. Salehi, S. L. Mortazavifar, M. Shahraki, and E. Abiri, *Opt. Eng.*, 2023, **62**, 035109.
[46] Z. Salamon, H. A. Macleod, and G. Tollin, *Biophys. J.*, 1997, **73**, 2791.
[47] Z. Salamon, G. Lindblom, L. Rilfors, K. Linde, and G. Tollin, *Biophys. J.*, 2000, **78**, 1400.
[48] Z. Salamon, M. F. Brown, and G. Tollin, *Trends Biochem. Sci.*, 1999, **24**, 213.
[49] Z. Salamon, G. Lindblom, and G. Tollin, *Biophys. J.*, 2003, **84**, 1796.
[50] S. Mahajna, M. Neumann, O. Eyal, and A. Shalabney, *J. Sens.*, 2016, **6**, 1898315.
[51] A. Shalabney and I. Abdulhalim, *Sens. Actuators A*, 2010, **159**, 24.